\documentclass[aps,eqsecnum,preprint,floats,epsf,epsfig,nofootinbib,letter]{revtex4}
\usepackage{epsfig}

\begin{document}
\def\be{\begin{eqnarray}}
\def\en{\end{eqnarray}}
\def\non{\nonumber}
\def\la{\langle}
\def\ra{\rangle}
\def\nc{N_c^{\rm eff}}
\def\vp{\varepsilon}
\def\drho{\bar\rho}
\def\deta{\bar\eta}
\def\CP{{\it CP}~}
\def\a{{\cal A}}
\def\B{{\cal B}}
\def\c{{\cal C}}
\def\d{{\cal D}}
\def\e{{\cal E}}
\def\p{{\cal P}}
\def\t{{\cal T}}
\def\up{\uparrow}
\def\dw{\downarrow}
\def\vma{{_{V-A}}}
\def\vpa{{_{V+A}}}
\def\smp{{_{S-P}}}
\def\spp{{_{S+P}}}
\def\J{{J/\psi}}
\def\ov{\overline}
\def\Lqcd{{\Lambda_{\rm QCD}}}
\def\pr{{Phys. Rev.}~}
\def\prl{{Phys. Rev. Lett.}~}
\def\pl{{Phys. Lett.}~}
\def\np{{Nucl. Phys.}~}
\def\zp{{Z. Phys.}~}
\def\lsim{ {\ \lower-1.2pt\vbox{\hbox{\rlap{$<$}\lower5pt\vbox{\hbox{$\sim$}
}}}\ } }
\def\gsim{ {\ \lower-1.2pt\vbox{\hbox{\rlap{$>$}\lower5pt\vbox{\hbox{$\sim$}
}}}\ } }

\font\el=cmbx10 scaled \magstep2{\obeylines\hfill May, 2007}

\vskip 1.5 cm

\centerline{\large\bf Charmless $B$ decays to a scalar meson and a
vector meson}
\bigskip
\centerline{\bf Hai-Yang Cheng$^{1}$, Chun-Khiang Chua$^{2}$ and
Kwei-Chou Yang$^{2}$}
\medskip
\centerline{$^1$ Institute of Physics, Academia Sinica}
\centerline{Taipei, Taiwan 115, Republic of China}
\medskip
\centerline{$^2$ Department of Physics, Chung Yuan Christian
University} \centerline{Chung-Li, Taiwan 320, Republic of China}
\bigskip
\bigskip
\bigskip
\bigskip
\centerline{\bf Abstract}
\bigskip
\small
 The hadronic charmless $B$ decays into a scalar meson and a
vector meson are studied within the framework of QCD
factorization. The main results are: (i) The decay rates for the
$f_0(980)K^{*-}$ and $f_0(980)\ov K^{*0}$ modes depend on the
$f_0-\sigma$ mixing angle $\theta$. The experimental measurements
can be accommodated for $\theta\approx 20^\circ$. (ii) If the
$a_0(980)$ is a $q\bar q$ bound state, the predicted branching
ratios for the channels $a_0^+\rho^-$ and $a_0^0\rho^-$ will be
very large, of order $30\times 10^{-6}$ and $23\times 10^{-6}$,
respectively. If the observation of or the experimental limit on
theses two modes is much smaller than the expectation of $\sim
25\times 10^{-6}$, this could hint at a four-quark nature for the
$a_0(980)$. (iii) For the $a_0(1450)$ channels,
$a_0^+(1450)\rho^-$ and $a_0^0(1450)\rho^-$ are found to have
branching ratios of order $16\times 10^{-6}$ and $22\times
10^{-6}$, respectively. A measurement of them at the predicted
level will favor the $q\bar q$ structure for the $a_0(1450)$. (iv)
Contrary to the naive expectation that $\Gamma(B^-\to
a_0^0\rho^-)\sim {1\over 2}\Gamma(\ov B^0\to a_0^+\rho^-)$, we
found that they have comparable rates due to additional
contributions  to the $a_0^0\rho^-$ mode from the $a_0^0$
emission. (v) The predicted central value of  $\B(\ov B^0\to\ov
K^{*0}_0(1430)\phi)$ is larger than experiment, though it can be
accommodated within theoretical errors. The decays $B\to
K^{*}_0(1430)\rho$ are expected to have rates substantially larger
than that of $B\to K^{*}_0(1430)\pi$ owing to the constructive
(destructive) interference between the $a_4$ and $a_6$ penguin
terms in the former (latter). Experimentally, it is thus important
to check if the $B\to K^{*}_0\rho$ modes are enhanced relative to
the corresponding $K_0^*\pi$ channels.

\pagebreak

\section{Introduction}
Recently we have studied the hadronic charmless $B$ decays into a
scalar meson and a pseudoscalar meson within the framework of QCD
factorization (QCDF) \cite{CCY}. It is known that the
identification of scalar mesons is difficult experimentally and
the underlying structure of scalar mesons is not well established
theoretically (for a review, see e.g.
\cite{Spanier,Godfrey,Close}). The experimental measurements of
$B\to SP$ will provide valuable information on the nature of the
even-parity mesons. For example, it was pointed out in \cite{CCY}
that the predicted $\ov B^0\to a_0^\pm(980)\pi^\mp$ and
$a_0^+(980)K^-$ rates exceed the current experimental limits,
favoring a four-quark nature for the $a_0(980)$. The decay $\ov
B^0\to \kappa^+ K^-$ also provides a nice ground for testing the
4-quark and 2-quark structure of the $\kappa$ (or $K_0^*(800)$)
meson. It can proceed through $W$-exchange and hence is quite
suppressed if the $\kappa$ is made of $q\bar q$ quarks, while it
receives a tree contribution if the $\kappa$ is predominately a
four-quark state. Hence, an observation of this channel at the
level of $\gsim 10^{-7}$ may imply a four-quark assignment for the
$\kappa$ \cite{CCY}

In this work we shall generalize our previous study to the decays
$B\to SV$ ($S$: scalar meson, $V$: vector meson), motivated by the
recent observation of the $\ov K_0^{*0}(1430)\phi$ and
$f_0(980)K^{*-}$ modes by BaBar:
 \be \label{eq:data}
 \B(\ov B^0\to \ov K_0^{*0}(1430)\phi) &=& (4.6\pm0.7\pm0.6)\times
 10^{-6}~[5], \non \\
 \B(B^-\to f_0(980)K^{*-};f_0(980)\to\pi^+\pi^-) &=& (5.2\pm1.2\pm0.5)\times
 10^{-6}~[6],  \non \\
 \B(\ov B^0\to f_0(980)\ov K^{*0};f_0(980)\to\pi^+\pi^-) &=& (2.6\pm0.6\pm0.9)\times 10^{-6}<4.3\times 10^{-6}~[6], \non \\
 \B(B^-\to f_0(980)\rho^-;f_0(980)\to\pi^+\pi^-) &<& 1.9\times 10^{-6} ~[7],  \non \\
  \B(\ov B^0\to f_0(980)\rho^0;f_0(980)\to\pi^+\pi^-) &<& 0.53\times 10^{-6}
 ~[8], \non \\
 \B(\ov B^0\to f_0(980)\omega;f_0(980)\to\pi^+\pi^-) &<& 1.5\times 10^{-6}
 ~[9].
 \en
Recently, the decay $\ov B^0\to \ov K_0^{*0}(1430)\phi$ has been
studied in \cite{Chen} within the framework of generalized
factorization in which the nonfactorizable effects are described
by the parameter $N_c^{\rm eff}$, the effective number of colors.
The result is sensitive to $N_c^{\rm eff}$. For example, the
branching ratio is predicted to be $(7.70,3.95,1.84)\times
10^{-6}$ for $N_c^{\rm eff}=(2,3,5)$. Hence, in the absence of
information for nonfactorizable effects, one cannot make a precise
prediction of the branching ratio. A QCDF calculation of this and
other modes will be presented in this work.

Since $B\to SP$ decays have been systematically explored in
\cite{CCY}, it is straightforward to generalize the study to the
$SV$ modes. In the sector of odd-parity mesons,  it is known that
the rates of the penguin-dominated modes $K^*\pi$ and $K\rho$ are
smaller than that of the corresponding $K\pi$ ones by a factor of
$\sim 2$. This can be understood as follows. In the factorization
approach, the penguin terms $a_6$ and $a_8$ are absent in the
decay amplitudes of $B\to K^*\pi$, while the effective Wilson
coefficients $a_4$ and $a_6$ contribute destructively to $B\to
K\rho$. In contrast, the tree-dominated $\rho\pi$ modes  have
rates larger than that of  $\pi\pi$ with the same charge
assignment due mainly to the fact that the $\rho$ meson has a
decay constant larger than the pion. We shall see in the present
work that the same analog is not always true for the scalar meson
sector. For example, we will show that the rates for $\ov
K_0^{*0}(1430)\rho^{-,0}$ are larger than that of $\ov
K_0^{*0}(1430)\pi^{-,0}$.

The layout of the present paper is as follows. In Sec. II we
introduce the input quantities relevant to the present work, such
as the decay constants, form factors and light-cone distribution
amplitudes. We then apply QCD factorization in Sec. III to study
$B\to SV$ decays. Results and discussions are presented in Sec.
IV. Sec. V contains our conclusions. The factorizable amplitudes
of various $B\to SV$ decays are summarized in Appendix A.


\section{Input quantities}
Since most of the essential input quantities are already discussed
in \cite{CCY}, here we shall just recapitulate the main inputs.

\subsection{Decay constants and form factors}
Decay constants of scalar and vector mesons are defined as
 \be \label{eq:decayc}
 && \la V(p)|\bar q_2\gamma_\mu q_1|0\ra=f_Vm_V\vp_\mu^*, \qquad
 \la S(p)|\bar q_2\gamma_\mu q_1|0\ra=f_S p_\mu,
 \qquad \la S|\bar q_2q_1|0\ra=m_S\bar f_S.
 \en
For vector mesons, there is an additional transverse decay
constant defined by
 \be
 \la V(p,\vp^*)|\bar q\sigma_{\mu\nu}q'|0\ra=f_V^\bot(p_\mu
 \vp_\nu^*-p_\nu\vp_\mu^*),
 \en
which is scale dependent. The neutral scalar mesons $\sigma$,
$f_0$ and $a_0^0$ cannot be produced via the vector current owing
to charge conjugation invariance or conservation of vector
current:
 \be
 f_{\sigma}=f_{f_0}=f_{a_0^0}=0.
 \en
For other scalar mesons, the vector decay constant $f_S$ and the
scale-dependent scalar decay constant $\bar f_S$ are related by
equations of motion
 \be \label{eq:EOM}
 \mu_Sf_S=\bar f_S, \qquad\quad{\rm with}~~\mu_S={m_S\over
 m_2(\mu)-m_1(\mu)},
 \en
where $m_{2}$ and $m_{1}$ are the running current quark masses and
$m_S$ is the scalar meson mass. For the neutral scalar mesons
$f_0$, $a_0^0$ and $\sigma$, $f_S$ vanishes, but the quantity
$\bar f_S=f_S\mu_S$ remains finite.

In \cite{CCY} we have applied the QCD sum rule method to estimate
various decay constants for scalar mesons which are summarized as
follows:
 \be \label{eq:decaycS}
 \bar f_{a_0(980)}(1~{\rm GeV})= (365\pm 20)~{\rm MeV},\ \ &
 \bar f_{a_0(980)}(2.1~{\rm GeV})= (450\pm 25)~{\rm MeV}, \non \\
 \bar f_{f_0(980)}(1~{\rm GeV})= (370\pm 20)~{\rm MeV},  \ \ &
 \bar f_{f_0(980)}(2.1~{\rm GeV})= (460\pm 25)~{\rm MeV}, \non \\
 \bar f_{a_0(1450)}(1~{\rm GeV})= (460\pm 50)~{\rm MeV},  \ \ &
 \bar f_{a_0(1450)}(2.1~{\rm GeV})= (570\pm 60)~{\rm MeV}, \non \\
 \bar f_{K_0^*(1430)}(1~{\rm GeV})= (445\pm 50)~{\rm MeV}, \ \ &
 \bar f_{K_0^*(1430)}(2.1~{\rm GeV})= (550\pm 60)~{\rm MeV}.
 \en
In \cite{CCY} we have considered two different scenarios for the
scalar mesons above 1 GeV, which will be briefly discussed in Sec.
IV. The above decay constants for the $a_0(1450)$ and
$K_0^*(1430)$ are obtained in scenario II. Using the running quark
masses given in Eq. (\ref{eq:quarkmass}) we obtain the
scale-independent decay constants:
 \be
 f_{a_0(980)^\pm}=1.0\,{\rm MeV}, \qquad  f_{a_0(1450)^\pm}=5.3\,{\rm MeV}, \qquad
 f_{K^*_0(1430)}=35.9\,{\rm MeV}.
 \en
For longitudinal and transverse decay constants of the vector
mesons, we use (in MeV)
 \be
 && f_\rho=216\pm3, \qquad f_\omega=187\pm5, \qquad~ f_{K^*}=220\pm5,
 \qquad~~ f_\phi=215\pm5, \non \\
 && f_\rho^\bot=165\pm9, \qquad f_\omega^\bot=151\pm9, \qquad f_{K^*}^\bot=185\pm10, \qquad
 f_\phi^\bot=186\pm9\,,
 \en
where the values of $f_V$ and $f_V^\bot$ are taken from
\cite{BallfV}.

Form factors for $B\to S,V$ transitions are defined by \cite{BSW}
 \be \label{eq:FF}
   \la V(p')|V_\mu|B(p)\ra &=& -{1\over
m_B+m_V}\,\epsilon_{\mu\nu\alpha \beta}\vp^{*\nu}P^\alpha
q^\beta  V^{PV}(q^2),   \non \\
 \la V(p')|A_\mu|B(p)\ra &=&  i\Big\{
(m_B+m_V)\vp^{*}_\mu A_1^{PV}(q^2)-{\vp^{*}\cdot P\over
m_B+m_V}\,P_\mu A_2^{PV}(q^2)    \non \\
&& -2m_V\,{\vp^{*}\cdot P\over
q^2}\,q_\mu\big[A_3^{PV}(q^2)-A_0^{PV}(q^2)\big]\Big\}, \non \\
\la S(p')|A_\mu|B(p)\ra &=& -i\Bigg[\left(P_\mu-{m_B^2-m_S^2\over
q^2}\,q_ \mu\right) F_1^{BS}(q^2)   +{m_B^2-m_S^2\over
q^2}q_\mu\,F_0^{BS}(q^2)\Bigg],
 \en
where $P_\mu=(p+p')_\mu$, $q_\mu=(p-p')_\mu$.  As shown in
\cite{CCH}, a factor of $(-i)$ is needed in $B\to S$ transition in
order for the $B\to S$ form factors to be positive. This also can
be checked from heavy quark symmetry \cite{CCH}.

Various form factors for $B\to S,V$ transitions have been
evaluated in the relativistic covariant light-front quark model
\cite{CCH}. In this model form factors are first calculated in the
spacelike region and their momentum dependence is fitted to a
3-parameter form
  \be \label{eq:FFpara}
 F(q^2)=\,{F(0)\over 1-a(q^2/m_{B}^2)+b(q^2/m_{B}^2)^2}\,.
 \en
The parameters $a$, $b$ and $F(0)$ are first determined in the
spacelike region. This parametrization is then analytically
continued to the timelike region to determine the physical form
factors at $q^2\geq 0$.  The results relevant for our purposes are
summarized in Table \ref{tab:FF}.   The form factors for $B$ to
$f_0(980)$ and $a_0(980)$ transitions are taken to be 0.25 at
$q^2=0$ \cite{CCY}.


\begin{table}[t]
\caption{Form factors for $B\to\rho,K^*,a_0(1450),K_0^*(1430)$
transitions obtained from the covariant light-front model
\cite{CCH}. } \label{tab:FF}
\begin{ruledtabular}
\begin{tabular}{| l c c c c | l c c c c |}
~~~$F$~~~~~~~~~
    & ~~$F(0)$
    & $F(q^2_{\rm max})$
    &$a$~~
    & $b$~~~~~
& ~~~ $F$~~~
    & $F(0)$~
    & $F(q^2_{\rm max})$~
    & $a$~~
    & $b$~~
 \\
    \hline
$F^{Ba_0(1450)}_1$
    & $0.26$
    & $0.68$
    & 1.57
    & 0.70
 &$F^{Ba_0(1450)}_0$
    & 0.26
    & 0.35
    & $0.55$
    & 0.03
    \\
$F^{BK^*_0(1430)}_1$
    & $0.26$
    & $0.70$
    & 1.52
    & 0.64
&$F^{BK^*_0(1430)}_0$
    & 0.26
    & 0.33
    & 0.44
    & 0.05
    \\
$V^{B\rho}$
    & $0.27$
    & $0.79$
    & 1.84
    & 1.28
&$A^{B\rho}_0$
    & 0.28
    & 0.76
    & 1.73
    & 1.20
    \\
$V^{BK^*}$
    & $0.31$
    & $0.96$
    & 1.79
    & 1.18
&$A^{BK^*}_0$
    & 0.31
    & 0.87
    & 1.68
    & 1.08
    \\
\end{tabular}
\end{ruledtabular}
\end{table}


We need to pay a special attention to the decay constants and form
factors for the $f_0(980)$. What is the quark structure of the
light scalar mesons below or near 1 GeV has been quite
controversial. In this work we shall consider the conventional
$q\bar q$ assignment for the light scalars $f_0(980)$ and
$a_0(980)$. In the naive quark model, the flavor wave functions of
the $f_0(980)$ and $\sigma(600)$ read
 \be
 \sigma={1\over \sqrt{2}}(u\bar u+d\bar d), \qquad\qquad f_0= s\bar
 s,
 \en
where the ideal mixing for $f_0$ and $\sigma$ has been assumed. In
this picture, $f_0(980)$ is purely an $s\bar s$ state. However,
there also exist some experimental evidences indicating that
$f_0(980)$ is not purely an $s\bar s$ state (see \cite{ChengDSP}
for details). Therefore, isoscalars $\sigma(600)$ and $f_0$ must
have a mixing
 \be
 |f_0(980)\ra = |s\bar s\ra\cos\theta+|n\bar n\ra\sin\theta,
 \qquad |\sigma(600)\ra = -|s\bar s\ra\sin\theta+|n\bar n\ra\cos\theta,
 \en
with $n\bar n\equiv (\bar uu+\bar dd)/\sqrt{2}$. Experimental
implications for the $f_0\!-\!\sigma$ mixing angle have been
discussed in detail in \cite{ChengDSP}.  In this work, we shall
use $\theta=20^\circ$, which is favored by the phenomenological
analysis of $B\to f_0K^*$ decays (see below). In the decay
amplitudes involving the $f_0(980)$ we will use the superscripts
$q=u,d,s$ to indicate that it is the $q$ quark content of the
$f_0(980)$ that gets involved in the interaction. For example,
$\bar f_{f_0}^s=\bar f_{f_0}\cos\theta$ and
$F_1^{Bf_0^u}=F_1^{Bf_0}\sin\theta/\sqrt{2}$.

\subsection{Distribution amplitudes}
The twist-2 light-cone distribution amplitude (LCDA) $\Phi_S(x)$
and twist-3 $\Phi_S^s(x)$ and $\Phi_S^\sigma(x)$ for the scalar
meson $S$ respect the normalization conditions
 \be \label{eq:wfnor}
 \int_0^1dx \Phi_S(x)=f_S, \qquad \int_0^1dx \Phi_S^s(x)=
 \int_0^1dx \Phi_S^\sigma(x)=\bar f_S.
 \en
In general, the twist-2 light-cone distribution amplitude $\Phi_S$
has the form
 \be \label{eq:twist2wf}
 \Phi_S(x,\mu)=f_S\,6x(1-x)\left[1+\mu_S\sum_{m=1}^\infty
 B_m(\mu)\,C_m^{3/2}(2x-1)\right],
 \en
where $B_m$ are Gegenbauer moments and $C_m^{3/2}$ are the
Gegenbauer polynomials.  For the neutral scalar mesons
$f_0,a_0^0,\sigma$, only odd Gegenbauer polynomials contribute. In
\cite{CCY} we have applied the QCD sum rules to determine the
Gegenbauer moments $B_1$ and $B_3$ (see Table
\ref{tab:momentscenario1}). For twist-3 LCDAs, we use
 \be \label{eq:twist3wf}
 \Phi^s_S(x)=\bar f_S , \qquad \Phi^\sigma_S(x)=\bar f_S\,
 6x(1-x).
 \en

\begin{table}[htb]
\caption{Gegenbauer moments $B_1$ and $B_3$ at the scales $\mu=1$
GeV and 2.1 GeV (shown in parentheses) obtained using the QCD sum
rule method \cite{CCY}. } \label{tab:momentscenario1}
\begin{ruledtabular}
\begin{tabular}{l rr}
 State & $B_1$ & $B_3$ \\ \hline
 $a_0(980)$
  & $-0.93\pm 0.10 (-0.64\pm 0.07)$ & $0.14\pm 0.08\ \ (0.08\pm 0.04)$      \\
 $f_0(980)$
  & $-0.78\pm0.08 (-0.54\pm 0.06)$ & $0.02\pm 0.07\ \ (0.01\pm 0.04)$      \\
 $a_0(1450)$
  & $-0.58\pm 0.12 (-0.40\pm 0.08)$ & $-0.49\pm 0.15 (-0.29\pm 0.09)$  \\
 $K_0^*(1430)$
  & $-0.57\pm 0.13 (-0.39\pm 0.09)$ & $-0.42\pm 0.22 (-0.25\pm 0.13)$
\end{tabular}
\end{ruledtabular}
\end{table}

For vector mesons, the normalization for the twist-2 function
$\Phi_V$ and the twist-3 function $\Phi_v$ is given by \cite{BN}
 \be \label{eq:Vwf}
 \int_0^1dx\Phi_V(x)=f_V, \qquad  \int^1_0 dx\Phi_v(x)=0,
 \en
where the definitions for $\Phi_v(x)$ can be found in \cite{BN}.
The general expressions of these LCDAs read
 \be
 \Phi_V(x,\mu)=6x(1-x)f_V\left[1+\sum_{n=1}^\infty
 \alpha_n^V(\mu)C_n^{3/2}(2x-1)\right],
 \en
and
 \be
 \Phi_v(x,\mu)=3f_V^\bot\left[2x-1+\sum_{n=1}^\infty
 \alpha_{n,\bot}^V(\mu)P_{n+1}(2x-1)\right],
 \en
where $P_n(x)$ are the Legendre polynomials. The Gegenbauer
moments $\alpha_n^V$ and $\alpha^V_{n,\bot}$ have been studied
using the QCD sum rule method. Here we employ the most recent
updated values evaluated at $\mu=1$ GeV \cite{Ball}
 \be
 && \alpha_1^{K^*}=0.03\pm0.02, \quad \alpha_{1,\bot}^{K^*}=0.04\pm0.03, \quad
 \alpha_2^{K^*}=0.11\pm0.09, \quad
 \alpha_{2,\bot}^{K^*}=0.10\pm0.08,\non \\
 &&  \alpha_2^{\rho,\omega}=0.15\pm0.07, \quad
 \alpha_{2,\bot}^{\rho,\omega}=0.14\pm0.06,\quad \alpha_2^\phi=0.18\pm0.08,
 \quad~~ \alpha_{2,\bot}^\phi=0.14\pm0.07,
 \en
and $\alpha_1^V=0$, $\alpha_{1,\bot}^V$=0 for
$V=\rho,\omega,\phi$.

As stressed in \cite{CCY}, it is most suitable to define the LCDAs
of scalar mesons including decay constants. However, in order to
make connections between $B\to SV$ and $B\to VV$ amplitudes, it is
more convenient to factor out the decay constants in the LCDAs and
put them back in the appropriate places. In the ensuing
discussions, we will use the LCDAs with the decay constants
$f_S,\bar f_S,f_V, f_V^\bot,f_P$ being factored out.

\section{$B\to SV$ decays in QCD factorization}

 We shall use the QCD factorization approach
\cite{BBNS,BN} to study the short-distance contributions to the
$B\to SV$ decays with $S=f_0(980),a_0(980),a_0(1450),K^*_0(1430)$,
and $V=\rho,K^*,\phi,\omega$. In QCD factorization, the
factorizable amplitudes of above-mentioned decays are summarized
in Appendix A. The effective parameters $a_i^p$ with $p=u,c$ in
Eq. (\ref{eq:SDAmp}) can be calculated in the QCD factorization
approach \cite{BBNS}. They are basically the Wilson coefficients
in conjunction with short-distance nonfactorizable corrections
such as vertex corrections and hard spectator interactions. In
general, they have the expressions \cite{BBNS,BN}
 \be \label{eq:ai}
 a_i^p(M_1M_2) &=& \left(c_i+{c_{i\pm1}\over N_c}\right)N_i(M_2)
  +{c_{i\pm1}\over N_c}\,{C_F\alpha_s\over
 4\pi}\Big[V_i(M_2)+{4\pi^2\over N_c}H_i(M_1M_2)\Big]+P_i^p(M_2),
 \en
where $i=1,\cdots,10$,  the upper (lower) signs apply when $i$ is
odd (even), $c_i$ are the Wilson coefficients,
$C_F=(N_c^2-1)/(2N_c)$ with $N_c=3$, $M_2$ is the emitted meson
and $M_1$ shares the same spectator quark with the $B$ meson. The
quantities $V_i(M_2)$ account for vertex corrections,
$H_i(M_1M_2)$ for hard spectator interactions with a hard gluon
exchange between the emitted meson and the spectator quark of the
$B$ meson and $P_i(M_2)$ for penguin contractions. The expression
of the quantities $N_i(M_2)$ reads
 \be
 N_i(M_2)=\cases{0, & $i=6,8$~and~$M_2=V$, \cr
                 1, & else. \cr}
 \en
Note that $N_i(M_2)$ vanishes for $i=6,8$ and $M_2=V$ owing to the
consequence of the second equation in Eq. (\ref{eq:Vwf}).

The vertex and penguin corrections for $SV$ final states have the
same expressions as those for $PP$ and $PV$ states and can be
found in \cite{BBNS,BN}. Using the general LCDA
 \be
 \Phi_M(x,\mu)=6x(1-x)\left[1+\sum_{n=1}^\infty
 \alpha_n^M(\mu)C_n^{3/2}(2x-1)\right]
 \en
and applying Eq. (37) in \cite{BN} for vertex corrections, we
obtain
 \be
 V_i(M) &=& 12\ln{m_b\over\mu}-18-{1\over
 2}-3i\pi+\left({11\over
 2}-3i\pi\right)\alpha_1^M-{21\over 20}\alpha_2^M+\left({79\over 36}-{2i\pi\over
 3}\right)\alpha_3^M+\cdots, \non \\
 \en
for $i=1-4,9,10$,
 \be
 V_i(M) &=& -12\ln{m_b\over\mu}+6-{1\over
 2}-3i\pi-\left({11\over
 2}-3i\pi\right)\alpha_1^M-{21\over 20}\alpha_2^M-\left({79\over 36}-{2i\pi\over
 3}\right)\alpha_3^M+\cdots, \non \\
 \en
for $i=5,7$ and
 \be
 V_i(M)=\cases{ -6 & for~$M=S$, \cr
                  9-6\pi i & for~$M=V$, }
 \en
for $i=6,8$ in the NDR scheme for $\gamma_5$. The expressions of
$V_i(M)$ up to the $\alpha_2^M$ term are the same as that in
\cite{BBNS}.

As for the hard spectator function $H$, it reads
 \be \label{eq:Hi}
 H_i(M_1M_2) &=& -{f_Bf_{M_1}\over D(M_1M_2)
}\int^1_0 {d\rho\over\rho}\, \Phi_B(\rho)\int^1_0 {d\xi\over
\bar\xi} \,\Phi_{M_2}(\xi)\int^1_0 {d\eta\over \deta}\left[\pm
\Phi_{M_1}(\eta)+ r_\chi^{M_1}\,{\bar\xi\over
\xi}\,\Phi_{m_1}(\eta)\right], \non \\
 \en
for $i=1-4,9,10$, where the upper sign is for $M_1=V$ and the
lower sign for $M_1=S$,
 \be \label{eq:Hi57}
 H_i(M_1M_2) &=& {f_Bf_{M_1}\over D(M_1M_2)}\int^1_0 {d\rho\over\rho}\,
\Phi_B(\rho)\int^1_0 {d\xi\over \xi} \,\Phi_{M_2}(\xi)\int^1_0
{d\eta\over \deta}\left[\pm \Phi_{M_1}(\eta)+
r_\chi^{M_1}\,{\xi\over
\bar\xi}\,\Phi_{m_1}(\eta)\right], \non \\
 \en
for $i=5,7$ and $H_i=0$ for $i=6,8$,  $\bar\xi\equiv 1-\xi$ and
$\bar\eta\equiv 1-\eta$, $\Phi_M$ ($\Phi_m$) is the twist-2
(twist-3) light-cone distribution amplitude of the meson $M$, and
 \be
 D(SV)=F_1^{BS}(0)m_B^2, \qquad
 D(VS)=A_0^{BV}(0)m_B^2.
 \en
The ratios $r_\chi^V$ and $r_\chi^S$ are defined as
 \be \label{eq:rchiS}
 r_\chi^V(\mu)={2m_V\over m_b(\mu)}\,{f_V^\bot(\mu)\over f_V}, \qquad\quad
 r_\chi^S(\mu)={2m_S\over m_b(\mu)}\,{\bar f_S(\mu)\over f_S}={2m_S^2\over
 m_b(\mu)(m_2(\mu)-m_1(\mu))}.
 \en
For the neutral scalars $\sigma$, $f_0$ and $a_0^0$, $r_\chi^S$
becomes divergent while $f_S$ vanishes.  In this case one needs to
express $f_Sr_\chi^S$ by $\bar f_S\bar r_\chi^S$ with
 \be
 \bar r_\chi^S(\mu)={2m_S\over m_b(\mu)}.
 \en

Weak annihilation contributions are described by the terms $b_i$,
and $b_{i,{\rm EW}}$ in Eq. (\ref{eq:SDAmp}) which have the
expressions
 \be \label{eq:bi}
 b_1 &=& {C_F\over N_c^2}c_1A_1^i, \qquad\quad b_3={C_F\over
 N_c^2}\left[c_3A_1^i+c_5(A_3^i+A_3^f)+N_cc_6A_3^f\right], \non \\
 b_2 &=& {C_F\over N_c^2}c_2A_1^i, \qquad\quad b_4={C_F\over
 N_c^2}\left[c_4A_1^i+c_6A_2^f\right], \non \\
 b_{\rm 3,EW} &=& {C_F\over
 N_c^2}\left[c_9A_1^{i}+c_7(A_3^{i}+A_3^{f})+N_cc_8A_3^{i}\right],
 \non \\
 b_{\rm 4,EW} &=& {C_F\over
 N_c^2}\left[c_{10}A_1^{i}+c_8A_2^{i}\right],
 \en
where the subscripts 1,2,3 of $A_n^{i,f}$ denote the annihilation
amplitudes induced from $(V-A)(V-A)$, $(V-A)(V+A)$ and
$(S-P)(S+P)$ operators, respectively, and the superscripts $i$ and
$f$ refer to gluon emission from the initial and final-state
quarks, respectively.  Their explicit expressions can be obtained
from $A_n^{i,f}(VV)$ for the $VV$ case \cite{BenekeVV} with the
replacements specified in Eq. (\ref{eq:replacementI}):
 \be \label{eq:ann}
 A_1^{i}&=& \int\cdots \cases{
 \left( \Phi_V(x)\Phi_S(y)\left[{1\over x(1-\bar xy)}+{1\over
 x\bar y^2}\right]+r_\chi^Vr_\chi^S\Phi_v(x)\Phi_S^s(y)\,{2\over
 x\bar y}\right); & for~$M_1M_2=VS$,  \cr
 \left( \Phi_S(x)\Phi_V(y)\left[{1\over x(1-\bar xy)}+{1\over
 x\bar y^2}\right]+r_\chi^Vr_\chi^S\Phi_S^s(x)\Phi_v(y)\,{2\over
 x\bar y}\right); & for~$M_1M_2=SV$, } \non  \\
 A_2^{i}&=& \int\cdots \cases{
 \left(\Phi_V(x)\Phi_S(y)\left[{1\over \bar y(1-\bar xy)}+{1\over
 x^2\bar y}\right]+r_\chi^Vr_\chi^S\Phi_v(x)\Phi_S^s(y)\,{2\over
 x\bar y}\right); & for~$M_1M_2=VS$,  \cr
 \left( \Phi_S(x)\Phi_V(y)\left[{1\over \bar y(1-\bar xy)}+{1\over
 x^2\bar y}\right]+r_\chi^Vr_\chi^S\Phi_S^s(x)\Phi_v(y)\,{2\over
 x\bar y}\right); & for~$M_1M_2=SV$, }  \non \\
 A_3^{i}&=& \int\cdots \cases{ \left( r_\chi^V\Phi_v(x)\Phi_S(y)
 \,{2\bar x\over x\bar y(1-\bar x y)}-r_\chi^S\Phi_V(x)\Phi_S^s
 (y)\,{2y\over x\bar y(1-\bar x y)}\right); & for~$M_1M_2=VS$, \cr
 \left( -r_\chi^S\Phi_S^s(x)\Phi_V(y)
 \,{2\bar x\over x\bar y(1-\bar x y)}+r_\chi^V\Phi_S(x)\Phi_v
 (y)\,{2y\over x\bar y(1-\bar xy)}\right); & for~$M_1M_2=SV$,}
 \non \\
 A_3^{f} &=& \int\cdots \cases{ \left(r_\chi^V
 \Phi_v(x)\Phi_S(y)\,{2(1+\bar y)\over x\bar y^2}+r_\chi^S
 \Phi_V(x)\Phi_S^s(y)\,{2(1+x)\over x^2\bar y}\right); &
 for~$M_1M_2=VS$, \cr
 \left( -r_\chi^S
 \Phi_S^s(x)\Phi_V(y)\,{2(1+\bar y)\over x\bar y^2}-r_\chi^V
 \Phi_S(x)\Phi_v(y)\,{2(1+x)\over x^2\bar y}\right); &
 for~$M_1M_2=SV$,} \non \\
  A_1^f &=& A_2^f=0,
 \en
where $\int\cdots=\pi\alpha_s \int^1_0dxdy$, $\bar x=1-x$ and
$\bar y=1-y$. Note that we have adopted the same convention as in
\cite{BN} that $M_1$ contains an antiquark from the weak vertex
with longitudinal fraction $\bar y$, while $M_2$ contains a quark
from the weak vertex with momentum fraction $x$.

Using the asymptotic distribution amplitudes for vector mesons and
keeping the LCDA of the scalar meson to the third Gegenbaur
polynomial in Eq. (\ref{eq:twist2wf}), the annihilation
contributions can be simplified to
 \be
 A_1^i(VS) &\approx& 6\pi\alpha_s\left\{3\mu_S \left[B_1(3X_A+4-\pi^2)
+B_3\left(10X_A+{23\over 18}-{10\over 3}\pi^2\right)\right]
-r_\chi^S r_\chi^V X_A(X_A-2)\right\}, \non \\
 A_2^i(VS) &\approx& 6\pi\alpha_s\left\{3\mu_S\left[B_1(X_A+29-3\pi^2)
+B_3\left(X_A+{2956\over 9}-{100\over 3}\pi^2\right)\right]
 -r_\chi^S r_\chi^V X_A(X_A-2)\right\}, \non \\
 A_3^i(VS) &\approx& 6\pi\alpha_s \Bigg\{-r_\chi^V \mu_S
 \left[9B_1\left(X_A^2-4X_A-4+\pi^2\right)+10B_3\left(3X_A^2-
19X_A+{61\over 6}+3\pi^2\right)\right]
 \non \\ &-& r_\chi^S\left(X_A^2-2X_A
 +{\pi^2\over 3}\right)\Bigg\}, \non \\
 A_3^f(VS) &\approx& 6\pi\alpha_s \left\{-3r_\chi^V \mu_S
 (X_A-2)\left[B_1(6X_A-11)+B_3\left(20X_A-{187\over 3}\right)\right]+r_\chi^SX_A(2X_A-1) \right\},
 \non \\
 \en
for $M_1M_2=VS$, and
 \be
 A_1^i(SV)=-A_2^i(VS),\qquad \qquad  A_2^i(SV)=-A_1^i(VS), \non \\
  A_3^i(SV)=A_3^i(VS), \qquad \qquad  A_3^f(SV)=-A_3^f(VS),
 \en
for $M_1M_2=SV$, where the endpoint divergence $X_A$ is defined in
Eq. (\ref{eq:XA}) below. As noticed in passing, for neutral
scalars $\sigma$, $f_0$ and $a_0^0$, one needs to express $f_S
r_\chi^S$ by $\bar f_S\bar r_\chi^S$ and $f_S\mu_S$ by $\bar f_S$.
Numerically, the dominant annihilation contribution arises from
the factorizable penguin-induced annihilation characterized by
$A_3^f$. Physically, this is because the penguin-induced
annihilation contribution is not subject to helicity suppression.

Although the parameters $a_i(i\neq 6,8)$ and $a_{6,8}r_\chi$ are
formally renormalization scale and $\gamma_5$ scheme independent,
in practice there exists some residual scale dependence in
$a_i(\mu)$ to finite order. To be specific, we shall evaluate the
vertex corrections to the decay amplitude at the scale
$\mu=m_b/2$. In contrast, as stressed in \cite{BBNS}, the hard
spectator and annihilation contributions should be evaluated at
the hard-collinear scale $\mu_h=\sqrt{\mu\Lambda_h}$ with
$\Lambda_h\approx 500 $ MeV. There is one more serious
complication about these contributions; that is, while QCD
factorization predictions are model independent in the
$m_b\to\infty$ limit, power corrections always involve troublesome
endpoint divergences. For example, the annihilation amplitude has
endpoint divergences even at twist-2 level and the hard spectator
scattering diagram at twist-3 order is power suppressed and posses
soft and collinear divergences arising from the soft spectator
quark. Since the treatment of endpoint divergences is model
dependent, subleading power corrections generally can be studied
only in a phenomenological way. We shall follow \cite{BBNS} to
parameterize the endpoint divergence $X_A\equiv\int^1_0 dx/\bar x$
in the annihilation diagram as
 \be \label{eq:XA}
 X_A=\ln\left({m_B\over \Lambda_h}\right)(1+\rho_A e^{i\phi_A}),
 \en
with the unknown real parameters $\rho_A$ and $\phi_A$. Likewise,
the endpoint divergence $X_H$ in the hard spectator contributions
can be parameterized in a similar manner.

\section{Results and discussions}

While it is widely believed that the $f_0(980)$ and the $a_0(980)$
are predominately four-quark states, in practice it is difficult
to make quantitative predictions on hadronic $B\to SV$ decays
based on the four-quark picture for light scalar mesons as it
involves not only the unknown form factors and decay constants
that are beyond the conventional quark model but also additional
nonfactorizable contributions that are difficult to estimate.
Hence, we shall assume the $q\bar q$ scenario for the $f_0(980)$
and the $a_0(980)$.

For $a_0(1450)V$ and $K^*_0(1430)$ channels, we have explored in
\cite{CCY} two possible scenarios for the scalar mesons above 1
GeV in the QCD sum rule method: (i) In scenario 1, we treat
$\kappa, a_0(980), f_0(980)$ as the lowest lying states, and
$K_0^*(1430), a_0(1450),f_0(1500)$ as the corresponding first
excited states, respectively, where we have assumed that
$f_0(980)$ and $f_0(1500)$ are dominated by the $\bar s s$
component and (ii) we assume in scenario 2 that $K_0^*(1430),
a_0(1450), f_0(1500)$ are the lowest lying resonances and the
corresponding first excited states lie between $(2.0\sim
2.3)$~GeV. Scenario 2 corresponds to the case that light scalar
mesons are four-quark bound states, while all scalar mesons are
made of two quarks in scenario 1. We found that the predicted
$a_0(980)K$ and $a_0^+(980)\pi^-$ rates in scenario 1 will be too
large compared to the current limits if the $a_0(980)$ is a bound
state of the conventional $q\bar q$ quarks. This means that the
scenario 2 in which the scalar mesons above 1 GeV are lowest lying
$q\bar q$ scalar state and the light scalar mesons are four-quark
states is preferable. Therefore, we shall use scenario 2 when
discussing $a_0(1450)$ and $K_0^*(1430)$ mesons.

\begin{table}[t]
\caption{Branching ratios (in units of $10^{-6}$) of $B$ decays to
final states containing a scalar meson and a vector meson. The
theoretical errors correspond to the uncertainties due to (i) the
Gegenbauer moments $B_{1,3}$, the scalar meson decay constants,
(ii) the heavy-to-light form factors and the strange quark mass,
and (iii) the power corrections due to weak annihilation and hard
spectator interactions, respectively. The predicted branching
ratios of $B\to f_0(980)K^*,f_0(980)\pi$ are for the $f_0-\sigma$
mixing angle $\theta=20^\circ$. For light scalar mesons $f_0(980)$
and $a_0(980)$ we have assumed the $q\bar q$ content for them. The
scalar mesons $a_0(1450)$ and $K_0^*(1450)$ are treated as the
lowest lying scalar states, corresponding to scenario 2 as
explained in Appendices B and C of \cite{CCY}. Experimental
results are taken from Eq. (\ref{eq:data}). We have assumed
$\B(f_0(980)\to \pi^+\pi^-)=0.50$ to obtain the experimental
branching ratios for $f_0(980)V$.} \label{tab:theoryBR}
\begin{ruledtabular}
\begin{tabular}{l r c |l  r c}
Mode & Theory & Expt & Mode & Theory & Expt  \\
\hline
 $B^-\to f_0(980)K^{*-}$ & $7.4^{+0.4+0.2+7.2}_{-0.4-0.2-2.9}$ & $10.4\pm2.6$ &
 $\ov B^0\to f_0(980)\ov K^{*0}$ & $6.4^{+0.4+0.3+7.0}_{-0.4-0.3-2.6}$  & $5.2\pm2.2<8.6$ \non \\
 $B^-\to f_0(980)\rho^-$ & $1.3^{+0.1+0.4+0.1}_{-0.1-0.3-0.1}$ & $<3.8$ &
 $\ov B^0\to f_0(980)\rho^0$ & $0.01^{+0.00+0.00+0.02}_{-0.00-0.00-0.01}$  & $<1.06$  \non \\
 & & &  $\ov B^0\to f_0(980)\omega$ & $0.06^{+0.02+0.00+0.02}_{-0.01-0.00-0.02}$  & $<3.0$  \non \\
 $B^-\to a_0^0(980)K^{*-}$ & $2.3^{+0.1+0.4+5.6}_{-0.1-0.4-1.3}$ &  & $\ov B^0\to a_0^+(980)K^{*-}$ &
 $3.9^{+0.1+0.7+3.5}_{-0.1-0.6-1.4}$ &   \non \\
 $B^-\to a_0^-(980)\ov K^{*0}$ & $6.1^{+0.2+1.1+4.7}_{-0.2-1.0-2.1}$ &  & $\ov B^0\to a_0^0(980)\ov K^{*0}$
 & $2.5^{+0.2+0.5+6.5}_{-0.2-0.4-1.8}$ &  \non \\
 $B^-\to a_0^0(980)\rho^-$ & $23.0^{+0.6+3.6+3.9}_{-0.5-3.3-2.6}$ &  & $\ov B^0\to a_0^+(980)\rho^-$
 & $29.5^{+2.4+5.9+6.0}_{-2.1-5.3-4.5}$ &   \non \\
 $B^-\to a_0^-(980)\rho^0$ & $3.0^{+0.8+0.1+2.3}_{-0.6-0.1-1.0}$ &  & $\ov B^0\to a_0^-(980)\rho^+$
 & $0.09^{+0.02+0.00+0.30}_{-0.02-0.00-0.07}$ &  \non \\
 $B^-\to a_0^-(980)\omega$ & $1.0^{+0.3+0.0+0.3}_{-0.2-0.0-0.3}$ & & $\ov B^0\to a_0^0(980)\rho^0$ &
 $1.9^{+0.3+0.0+0.2}_{-0.3-0.0-0.2}$ & \non \\
 & & &  $\ov B^0\to a_0^0(980)\omega$ &
 $1.3^{+0.2+0.0+0.1}_{-0.2-0.0-0.1}$ & \non \\
 $B^-\to a_0^0(1450)K^{*-}$ & $2.2^{+0.2+0.4+12.1}_{-0.2-0.3-~1.3}$ &  & $\ov B^0\to a_0^+(1450)K^{*-}$ &
 $4.7^{+0.3+0.7+19.1}_{-0.3-0.7-~2.3}$ &  \non \\
 $B^-\to a_0^-(1450)\ov K^{*0}$ & $7.8^{+0.5+1.2+23.6}_{-0.5-1.1-~4.6}$ &  & $\ov B^0\to a_0^0(1450)\ov K^{*0}$
 & $2.5^{+0.4+0.5+13.6}_{-0.3-0.4-~2.2}$ &  \non \\
 $B^-\to a_0^0(1450)\rho^-$ & $21.7^{+0.2+3.4+4.2}_{-0.2-3.1-3.9}$ & & $\ov B^0\to a_0^+(1450)\rho^-$
 & $16.0^{+2.8+4.3+20.8}_{-2.4-3.8-10.4}$ &  \non \\
 $B^-\to a_0^-(1450)\rho^0$ & $5.3^{+1.6+0.1+7.6}_{-1.3-0.1-2.4}$ &  & $\ov B^0\to a_0^-(1450)\rho^+$
 & $2.2^{+1.0+0.0+1.3}_{-0.7-0.0-0.5}$ &  \non \\
 $B^-\to a_0^-(1450)\omega$ & $1.7^{+0.6+0.0+1.3}_{-0.5-0.0-1.0}$ & &  $\ov B^0\to a_0^0(1450)\rho^0$ &
 $3.1^{+0.7+0.0+1.9}_{-0.6-0.1-1.2}$ & \non \\
 & & &  $\ov B^0\to a_0^0(1450)\omega$ &
 $1.8^{+0.5+0.0+1.3}_{-0.4-0.0-0.8}$ & \non \\
 $B^-\to K^{*-}_0(1430)\phi$ &  $16.7^{+6.1+1.6+52.1}_{-4.6-1.6-10.1}$ &
 & $\ov B^0\to \ov K^{*0}_0(1430)\phi$ &
 $16.4^{+6.1+1.6+51.6}_{-4.6-1.5-10.1}$& $4.6\pm0.9$ \non \\
 $B^-\to \ov K^{*0}_0(1430)\rho^-$ & $66.2^{+25.0+2.8+70.8}_{-19.5-2.4-26.3}$ & $$ &
 $\ov B^0\to K^{*-}_0(1430)\rho^+$  & $51.0^{+16.1+1.4+68.6}_{-13.1-1.2-23.8}$ & $$ \non \\
 $B^-\to K^{*-}_0(1430)\rho^0$ & $21.0^{+7.3+1.2+29.4}_{-5.9-1.1-10.1}$ &  &
 $\ov B^0\to \ov K^{*0}_0(1430)\rho^0$  & $36.8^{+14.3+0.9+23.4}_{-11.0-0.7-~9.1}$ &
 $$  \non \\
 $B^-\to K^{*-}_0(1430)\omega$ & $16.1^{+4.9+0.7+22.5}_{-4.0-0.6-~8.3}$ &  &
 $\ov B^0\to \ov K^{*0}_0(1430)\omega$  & $15.6^{+4.4+1.0+14.6}_{-3.7-0.8-~5.3}$ &
 $$  \non \\
\end{tabular}
\end{ruledtabular}
\end{table}

The calculated results for the branching ratios of $B\to SV$ are
shown in Table \ref{tab:theoryBR}. In the table we have included
theoretical errors arising from the uncertainties in the
Gegenbauer moments $B_{1,3}$ (cf. Table
\ref{tab:momentscenario1}), the scalar meson decay constant $f_S$
or $\bar f_S$ (see Eq. (\ref{eq:decaycS})), the form factors
$F^{BP,BS}$, the quark masses and the power corrections from weak
annihilation and hard spectator interactions characterized by the
parameters $X_A$ and $X_H$, respectively. For form factors we
assign their uncertainties to be $\delta F^{BP,BS}(0)=\pm 0.03$,
for example, $F_0^{BK}(0)=0.35\pm0.03$ and
$F_0^{BK^*_0}(0)=0.26\pm0.03$. The strange quark mass is taken to
be $m_s(2.1\,{\rm GeV})=90\pm20$~MeV. For the quantities $X_A$ and
$X_H$ we adopt the form (\ref{eq:XA}) with $\rho_{A,H}\leq 0.5$
and arbitrary strong phases $\phi_{A,H}$. Note that the central
values (or ``default" results) correspond to $\rho_{A,H}=0$ and
$\phi_{A,H}=0$.

\subsubsection{$B\to f_0(980)K^*$ and $a_0(980,1450)K^*$ decays}

The penguin-dominated $B\to f_0(980)K^*$ decay receives three
distinct types of factorizable contributions: one from the $K^*$
emission, one from the $f_0$ emission with the $s\bar s$ content,
and the other from the $f_0$ emission with the $n\bar n$
component. \footnote{In our previous work for $B\to SP$ decays
\cite{CCY}, we did not take into account the contributions from
the $f_0$ or the neutral $a_0$ emission induced from the
four-quark operators other than $O_6$ and $O_8$ (see also
\cite{Luf0}). Corrections will be published elsewhere.}
In the expression of $B\to f_0K^*$ decay amplitudes given in Eq.
(\ref{eq:SDAmp}), the superscript $u$ of the form factor
$F_0^{Bf_0^u}$ reminds us that it is the $u$ quark component of
$f_0$ involved in the form factor transition. In contrast, the
superscript $q$ of the decay constant $\bar f_{f_0}^q$ indicates
that it is the $q\bar q$ quark content of $f_0$ responsible for
the penguin contribution under consideration. Note that except for
the penguin operators $O_6$ and $O_8$, the $f_0$ emission
amplitudes induced from other four-quark operators contain a
vanishing $f_0$ decay constant. However, it is compensated by the
$\mu_S$ term in the twist-2 LCDA of the scalar meson so that the
combination  $f_{f_0}\,\mu_{f_0}=\bar f_{f_0}$ becomes finite.

\begin{figure}[t]
\vspace{0cm} \centerline{\psfig{figure=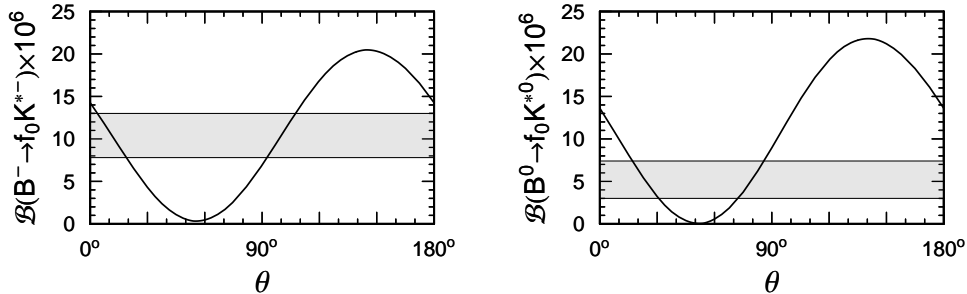,width=14cm}}
    \caption[]{\small Branching ratios of $B^-\to f_0(980)K^{*-}$ and $B^0\to f_0(980)K^{*0}$ versus the
    mixing angle $\theta$ of strange and nonstrange components of
    $f_0(980)$. For simplicity, theoretical errors  are not
    taken into account.
    The horizontal band
    shows the experimentally allowed region with one sigma error.}
    \label{fig:f0KV}
\end{figure}

In the extreme case that the $f_0(980)$ is made of $\bar ss$
quarks or $\bar nn$ quarks, the branching ratio of $B^-\to
f_0(980)K^{*-}$ is given by
 \be
 \B(B^-\to f_0(980)K^{*-})=\cases{(14.3^{+0.5+0.0+17.1}_{-0.4-0.0-~6.3})\times 10^{-6}; &
 for~$f_0(980)=\bar ss$, \cr  (6.9^{+1.1+0.6+12.5}_{-1.0-0.6-~3.9})\times 10^{-6}; &
 for~$f_0(980)=\bar nn$.}
 \en
In general, $\B(B\to f_0(980)K^{*})$ depends on the mixing angle
$\theta$ of strange and nonstrange components of the $f_0(980)$.
We found that the experimental data can be accommodated with
$\theta$ being in the vicinity of  $20^\circ$ (see Fig.
\ref{fig:f0KV}). The charged and neutral modes of $f_0(980)\ov
K^{*}$ are expected to have similar rates, while experimentally
their central values differ by a factor of 2. This discrepancy
needs to be clarified by the future improved measurements.

In order to compare theory with experiment for $B\to f_0(980)K^*$,
we need an input for $\B(f_0(980)\to \pi^+\pi^-)$. To do this, we
shall use the BES measurement \cite{BES}
 \be
 {\Gamma(f_0(980)\to \pi\pi)\over \Gamma(f_0(980)\to \pi\pi)+\Gamma(f_0(980)\to K\ov
K)}=0.75^{+0.11}_{-0.13}\,.
 \en
Assuming that the dominance of the $f_0(980)$ width by $\pi\pi$
and $K\ov K$ and applying isospin relation, we obtain
 \be
 \B(f_0(980)\to \pi^+\pi^-)=0.50^{+0.07}_{-0.09}\,, \qquad \B(f_0(980)\to
 K^+K^-)=0.125^{+0.018}_{-0.022}\,.
 \en
Hence, we use $\B(f_0(980)\to \pi^+\pi^-)=0.50$ to determine the
absolute branching ratio for $B\to f_0(980)K^*$.

For $a_0K^*$ decays, they have similar branching ratios as the
corresponding $a_0K$ channels \cite{CCY}.

\subsubsection{$B\to f_0(980)\rho$ and $a_0(980,1450)\rho$ decays}

The tree dominated decays $B\to a_0(980)\rho,f_0(980)\rho$ are
governed by the $B\to a_0$ and $B\to f_0^u$ transition form
factors, respectively. The $f_0\rho$ rate is rather small because
of the small $u\bar u$ component in the $f_0(980)$ and the
destructive interference between $a_4$ and $a_6$ penguin terms.
The $f_0\rho^0$ and $f_0\omega$ modes are suppressed relative to
$f_0\rho^-$ by at least a factor of ${1\over 2}|a_2/a_1|^2$.

The decay $\ov B^0\to a_0^+\rho^-$ has a rate much larger than the
$a_0^-\rho^+$ one because the factorizable amplitude of the former
(latter) is proportional to $f_\rho$ ($f_a$) and the decay
constant of the charged $a_0$ is very small. We also notice that
the predicted $a_0\rho^-$ rates are much larger than that of
$a_0\pi^-$ for two reasons. First of all, the $\rho$ meson decay
constant is bigger than that of the pion, $f_\rho\gg f_\pi$.
Second, the destructive interference between the $a_4$ and $a_6$
penguin terms is less severe for $a_0\rho$ as $r_\chi^\pi\sim
2.4\,r_\chi^\rho$. Contrary to the naive anticipation that
$\Gamma(B^-\to a_0^0\rho^-)\sim {1\over 2}\Gamma(\ov B^0\to
a_0^+\rho^-)$, we found that they have comparable rates due to
additional contributions  to the $a_0^0\rho^-$ mode from the
$a_0^0$ emission.

In the sector of the $a_0^0(980)\pi$ channels, we have argued
before that the fact that the experimental limits for the
$a_0^0\pi$ and $a_0^0K$  modes are smaller than the theoretical
expectations favors a four-quark nature for the $a_0(980)$
\cite{CCY}. Here we also suggest that if the observation of or the
experimental limit on the decay mode $a_0^+(980)\rho^-$
($a_0^0(980)\rho^-$)  is much smaller than the expectation of
$25\times 10^{-6}$, this could indicate a four-quark structure for
the $a_0(980)$.

Recently, the isovector scalar meson $a_0(1450)$ has been
confirmed to be a conventional $q\bar{q}$ meson in lattice
calculations~\cite{Mathur,lw00,bde02,kun04,pdi04}. Hence, the
calculations for the $a_0(1450)$ channels should be more
trustworthy. Our results indicate that $a_0^+(1450)\rho^-$ and
$a_0^0(1450)\rho^-$ have large branching ratios, of order
$16\times 10^{-6}$ and $22\times 10^{-6}$, respectively. A
measurement of them at the predicted level will reinforce the
$q\bar q$ nature for the $a_0(1450)$.

\subsubsection{$B\to K_0^*(1430)\phi$ and $K_0^*(1430)\rho$ decays}

For $K_0^*(1430)\phi$ channels, the central value of the predicted
$\B(\ov B^0\to \ov
K^*_0(1430)\phi)=(16.4^{+6.1+1.6+51.6}_{-4.6-1.5-10.1})\times
10^{-6}$ is larger than the experimental value of
$(4.6\pm0.9)\times 10^{-6}$, though they are consistent within
theoretical uncertainties. This mode was measured by BaBar
\cite{BaBarKstphi} using the LASS parametrization to describe the
$(K\pi)_0^{*0}$ amplitude. However, as commented in
\cite{BelleKpipi}, while this approach is experimentally
motivated, the use of the LASS parametrization is limited to the
elastic region of $M(K\pi)\lsim 2.0$ GeV, and an additional
amplitude is still required for a satisfactory description of the
data. Therefore, it will be interesting to see the Belle
measurement for $K_0^*(1430)\phi$ modes.

Theoretically, the $K^*_0(1430)\rho$ rates are expected to be
substantially larger than that of the $K^*_0(1430)\pi$ ones since
the penguins terms $a_4$ and $a_6$ contribute constructively to
the former and destructively to the latter. However, as shown in
\cite{CCY}, our predicted central values for the branching ratios
of $\ov K_0^{*0}\pi^-$ and $K_0^{*-}\pi^+$ are too small by a
factor $3\sim 4$ compared to experiment. \footnote{Recently, the
authors of \cite{Lu} claimed that the decay rates for the $\ov
K_0^{*0}\pi^-$ and $K_0^{*-}\pi^+$ modes can be accommodated in
the pQCD approach. It is not clear to us what is the underlying
reason for the discrepancy between our work and \cite{Lu}.
However, we have just performed a systematical study of charmless
3-body $B$ decays based on a simply generalized factorization
approach \cite{CCS}. We consider the weak process $B\to
K_0^*(1430)\pi$ followed by the strong decay $K_0^*\to K\pi$ and
reach the same conclusion as \cite{CCY}, namely, the predicted
$\ov K_0^{*0}\pi^-$ and $K_0^{*-}\pi^+$ rates are too small
compared to the data.}
It appears that one needs sizable weak annihilation in order to
accommodate the $K_0^*\pi$ data. In this work, we found large
rates for $\ov K_0^{*0}\rho^{-,0}$ and $K_0^{*-}\rho^+$ even in
the absence of weak annihilation contributions. Experimentally, it
should be relatively easy to search for those $K^*_0(1430)\rho$
modes to see if they are enhanced relative to their counterparts
in the $K^*_0\pi$ sector. The branching ratios for the
$K_0^*(1430)\omega$ modes are predicted to be of order $1.5\times
10^{-5}$.

\section{Conclusions}
We have studied the hadronic charmless $B$ decays into a scalar
meson and a vector meson within the framework of QCD
factorization. The main results are:

\begin{itemize}

\item The decay rates for the $f_0(980)K^{*-}$ and $f_0(980)\ov
K^{*0}$ modes depend on the mixing angle $\theta$ of strange and
nonstrange components of the $f_0(980)$. The experimental
measurements can be accommodated for $\theta\approx 20^\circ$.

\item  If the $a_0(980)$ is a $q\bar q$ bound state, the predicted
branching ratios for the channels $a_0^+\rho^-$ and $a_0^0\rho^-$
will be very large, of order $30\times 10^{-6}$ and $23\times
10^{-6}$, respectively. If the observation of or the experimental
limit on theses two modes is much smaller than the expectation of
$\sim 25\times 10^{-6}$, this could hint at a four-quark nature
for the $a_0(980)$.

\item For the $a_0(1450)$ channels, $a_0^+(1450)\rho^-$ and
$a_0^0(1450)\rho^-$ are found to have branching ratios of order
$16\times 10^{-6}$ and $22\times 10^{-6}$, respectively. An
observation of them at the predicted level will favor the $q\bar
q$ structure for the $a_0(1450)$.

\item Contrary to the naive expectation that $\Gamma(B^-\to
a_0^0\rho^-)\sim {1\over 2}\Gamma(\ov B^0\to a_0^+\rho^-)$, we
found that they have comparable rates due to additional
contributions  to the $a_0^0\rho^-$ mode from the $a_0^0$
emission.

\item  The predicted central value of $\B(\ov B^0\to\ov
K^{*0}_0(1430)\phi)$ is somewhat larger than experiment, though it
can be accommodated within theoretical errors. The decays $B\to
K^{*}_0(1430)\rho$ are expected to have rates substantially larger
than that of $B\to K^{*}_0(1430)\pi$ owing to the constructive
(destructive) interference between the $a_4$ and $a_6$ penguin
terms in the former (latter). Experimentally, it is thus important
to check if the $B\to K^{*}_0\rho$ modes are enhanced relative to
their counterparts in the $K^*_0\pi$ sector. The branching ratios
for the $K_0^*(1430)\omega$ modes are predicted to be of order
$1.5\times 10^{-5}$.

\end{itemize}

\vskip 2.5cm \acknowledgments

This research was supported in part by the National Science
Council of R.O.C. under Grant Nos. NSC95-2112-M-001-013,
NSC95-2112-M-033-001, and NSC95-2112-M-033-013.

\newpage
\appendix
\section{}

The $B\to SV$ ($VS$) decay amplitudes can be either evaluated
directly or obtained readily from $B\to VV$ amplitudes with the
replacements:
 \be \label{eq:replacementI}
 \quad \Phi_V(x)\to \Phi_S(x), \qquad \Phi_v(x)\to \Phi_S^s(x),
 \qquad f_V\to f_S, \qquad f_V^\bot\to -\bar f_S, \qquad r^V_\chi\to
 -r_\chi^S.
 \en
As stressed in the main text, we use the LCDAs with the decay
constants being factored out. Since the $VV$ channels have been
studied in details in \cite{BenekeVV}, we may use them to obtain
the $B\to SV$ amplitudes. In \cite{BenekeVV}, the factorizable
longitudinal $B\to VV$ amplitude reads (apart from the effective
Wilson coefficients)
 \be
A_{V_1V_2} &=&
i{G_F\over\sqrt{2}}\,2f_{V_2}A_0^{BV_1}(m_{V_2}^2)m_Bp_c,
 \en
where use has been made of the replacement $m_V\vp^*\cdot p_B\to
m_Bp_c$ with $p_c$ being the c.m. momentum. Since the definitions
for the decay constant $f_V$ and the form factor $A_0$ in
\cite{BenekeVV}
 \be
 \la V|V_\mu|0\ra=-if_Vm_V\vp^*_\mu, \qquad  \la V(p')|A_\mu|B(p)\ra =
2m_V\,{\vp^{*}\cdot P\over q^2}\,q_\mu A_0^{PV}(q^2)+\cdots
 \en
are different from ours [see Eqs. (\ref{eq:decayc}) and
(\ref{eq:FF})], the replacements (\ref{eq:replacementI}) need to
be modified accordingly. The $B\to VS$ amplitude is obtained from
the replacements:
 \be \label{eq:replacementVS}
 f_{V_1}\to if_V, \qquad f_{V_2}\to if_S, \qquad A_0^{BV_1} \to
 iA_0^{BV}, \qquad r_\chi^{V_2}\to -r_\chi^S.
 \en
For $B\to SV$ amplitudes, the replacements are
 \be \label{eq:replacementSV}
 f_{V_1}\to if_S, \qquad f_{V_2}\to if_V, \qquad A_0^{BV_1} \to
 -iF_1^{BS}, \qquad r_\chi^{V_2}\to r_\chi^V.
 \en
From (\ref{eq:replacementVS}) and (\ref{eq:replacementSV}) we
obtain the factorizable $B\to SV$ and $VS$ amplitudes
 \be
  A_{M_1M_2} &=& i{G_F\over\sqrt{2}}\cases{
 2f_VF_1^{BS}(m_V^2)m_Bp_c;  & for~$M_1M_2=SV$, \cr
 -2f_SA_0^{BV}(m_S^2)m_Bp_c; & for~$M_1M_2=VS$.}
 \en

The coefficients of the flavor operators $\alpha_i^p$ for $SV$ can
be obtained from the $VV$ case \cite{BenekeVV} and they read
 \be \label{eq:replacementII}
 \alpha_1(M_1,M_2) &=& a_1(M_1,M_2),\qquad \alpha_2(M_1M_2)=a_2(M_1M_2), \non \\
 \alpha_3^p(M_1M_2) &=& a_3^p(M_1M_2)+a_5^p(M_1M_2), \non \\
 \alpha_{\rm 3,EW}^p(M_1M_2) &=& a_9^p(M_1M_2)+a_7^p(M_1M_2), \non \\
 \alpha_4^p(M_1M_2) &=& \cases{ a_4^p(M_1M_2)-r_\chi^V
 a_6^p(M_1M_2); & for~$M_1M_2=SV$, \cr a_4^p(M_1M_2)+r_\chi^S
 a_6^p(M_1M_2); & for~$M_1M_2=VS$, } \non \\
 \alpha_{\rm 4,EW}^p(M_1M_2) &=& \cases{ a_{10}^p(M_1M_2)-r_\chi^V
 a_8^p(M_1M_2); & for~$M_1M_2=SV$, \cr a_{10}^p(M_1M_2)+r_\chi^S
 a_8^p(M_1M_2); & for~$M_1M_2=VS$. }
 \en

Applying the replacement (\ref{eq:replacementI}) and Eq.
(\ref{eq:replacementII}) to the $B\to VV$  amplitudes in
\cite{BenekeVV}, we obtain the following the factorizable
amplitudes of the decays $B\to
(f_0,a_0)K^*,~f_0(\rho,\omega),~a_0(\rho,\omega),~a_0K^*,~K_0^*(\phi,\rho,\omega)$:
 \be \label{eq:SDAmp}
A(B^- \to f_0 K^{*-} ) &=&
i\frac{G_F}{\sqrt{2}}\sum_{p=u,c}\lambda_p^{(s)}
 \Bigg\{ \left(a_1 \delta^p_u+a_4^p-r_\chi^{K^*}(a_6^p+a_8^p)
 +a_{10}^p \right)_{f_0^u K^*}
 2f_{K^*}F_1^{Bf_0^u}(m_{K^*}^2)m_Bp_c \non \\
 &-&\left(\bar a_3+\bar a_4^p+\bar a_5+(a_6^p-{1\over 2}a_8^p)
 \bar r_\chi^{f_0}-{1\over 2}(\bar a_7+\bar a_9
 +\bar a_{10}^p)\right)_{K^*f_0^s}2\bar f^s_{f_0}\,A_0^{BK^*}
 (m^2_{f_0})m_Bp_c\non \\
 &-&\left(\bar a_2\delta^p_u+2(\bar a_3+\bar a_5)+{1\over 2}(\bar a_7
 +\bar a_9)\right)_{K^*f_0^u}2\bar f^u_{f_0}\,A_0^{BK^*}
 (m^2_{f_0})m_Bp_c\non \\
 &-& f_Bf_{K^*}\bigg[\bar f_{f_0^u}\big(\bar b_2\delta_u^p+\bar b_3
 +\bar b_{\rm 3,EW}\big)_{f_0^uK^*} +\bar f_{f_0^s}\big(\bar b_2\delta_u^p+\bar b_3
 +\bar b_{\rm 3,EW}\big)_{K^*f_0^s}\bigg] \Bigg\}, \non \\
A(\ov B^0 \to f_0\ov K^{*0} ) &=&
i\frac{G_F}{\sqrt{2}}\sum_{p=u,c}\lambda_p^{(s)}
 \Bigg\{ \left(a_4^p-r_\chi^{K^*}(a_6^p-{1\over 2}a_8^p)
 -{1\over 2}a_{10}^p \right)_{f_0^dK^*}
 2f_{K^*}F_1^{Bf_0^u}(m_{K^*}^2)m_Bp_c \non \\
 &-&\left(\bar a_3+\bar a_4^p+\bar a_5+(a_6^p-{1\over 2}a_8^p)\bar r_\chi^{f_0}
 -{1\over 2}(\bar a_7+\bar a_9
 +\bar a_{10}^p)\right)_{K^*f_0^s}2\bar f^s_{f_0}\,A_0^{BK^*}
 (m^2_{f_0})m_Bp_c\non \\
 &-&\left(\bar a_2\delta^p_u+2(\bar a_3+\bar a_5)+{1\over 2}(\bar a_7
 +\bar a_9)\right)_{K^*f_0^u}2\bar f^u_{f_0}\,A_0^{BK^*}
 (m^2_{f_0})m_Bp_c\non \\
  &-& f_Bf_{K^*}\bigg[\bar f_{f_0^d}\big(\bar b_3
 -{1\over 2}\bar b_{\rm 3,EW}\big)_{f_0^dK^*} +f_{f_0^s}\big(\bar b_3
 -{1\over 2}\bar b_{\rm 3,EW}\big)_{K^*f_0^s}\bigg] \Bigg\}, \non \\
A(B^- \to a_0^0 K^{*-} ) &=&
i\frac{G_F}{{2}}\sum_{p=u,c}\lambda_p^{(s)}
 \Bigg\{ \left(a_1 \delta^p_u+a_4^p-r_\chi^{K^*}(a_6^p+a_8^p)
 +a_{10}^p \right)_{a_0 K^*}
  2f_{K^*}F_1^{Ba_0}(m_{K^*}^2)m_Bp_c \non \\
 &-&\left(\bar a_2\delta_u^p\right)_{K^*a_0}2\bar f_{a_0}A_0^{BK^*}(m_{a_0}^2)m_Bp_c
 -f_Bf_{K^*}\bar f_{a_0}\big(\bar b_2\delta_u^p+\bar b_3
 +\bar b_{\rm 3,EW}\big)_{a_0K^*}\Bigg\}, \non \\
A(B^- \to a_0^- \ov K^{*0} ) &=&
i\frac{G_F}{\sqrt{2}}\sum_{p=u,c}\lambda_p^{(s)}
 \Bigg\{ \left(a_4^p-r_\chi^{K^*}(a_6^p-{1\over 2}a_8^p)
 -{1\over 2}a_{10}^p \right)_{a_0 K^*} \non \\
&\times&  2f_{K^*}F_1^{Ba_0}(m_{K^*}^2)m_Bp_c
 - f_Bf_{K^*}f_{a_0}\big(b_2\delta_u^p+b_3
 +b_{\rm 3,EW}\big)_{a_0K^*} \Bigg\}, \non \\
A(\ov B^0 \to a_0^+ K^{*-} ) &=&
i\frac{G_F}{\sqrt{2}}\sum_{p=u,c}\lambda_p^{(s)}
 \Bigg\{ \left(a_1\delta_u^p+a_4^p-r_\chi^{K^*}(a_6^p+a_8^p)
 +a_{10}^p \right)_{a_0K^*} \non \\
&\times & 2f_{K^*}F_1^{Ba_0}(m_{K^*}^2)m_Bp_c
  - f_Bf_{K^*}f_{a_0}\big(b_3
 -{1\over 2}b_{\rm 3,EW}\big)_{a_0K^*}\bigg] \Bigg\}, \non \\
A(\ov B^0 \to a_0^0\ov K^{*0} ) &=&
 i\frac{G_F}{{2}}\sum_{p=u,c}\lambda_p^{(s)}
 \Bigg\{ -\left(a_4^p-r_\chi^{K^*}(a_6^p-{1\over 2}a_8^p)
 -{1\over 2}a_{10}^p \right)_{a_0K^*}
 2f_{K^*}F_1^{Ba_0}(m_{K^*}^2)m_Bp_c \non \\
   &-& \left(\bar a_2\delta_u^p\right)_{K^*a_0}2\bar f_{a_0}A_0^{BK^*}(m_{a_0}^2)m_Bp_c
  -f_Bf_{K^*}\bar f_{a_0}\big(-\bar b_3
 +{1\over 2}\bar b_{\rm 3,EW}\big)_{a_0K^*} \Bigg\}, \non \\
A(B^- \to f_0 \rho^- ) &=&
i\frac{G_F}{\sqrt{2}}\sum_{p=u,c}\lambda_p^{(d)}
 \Bigg\{ \left(a_1 \delta^p_u+a_4^p-r_\chi^\rho (a_6^p+a_8^p)
 +a_{10}^p \right)_{f_0^u \rho} \non \\
 &\times& 2f_\rho F_1^{Bf_0^u}(m_\rho^2)m_Bp_c
 - \Big(\bar a_2\delta_u^p+2(\bar a_3+\bar a_5)+\bar a_4+(a_6^p-{1\over
 2}a_8^p)\bar r_\chi^{f_0} \non \\
 &+& {1\over 2}(\bar a_7+\bar a_9-\bar a_{10})\Big)_{\rho f_0^u}2\bar f^u_{f_0}\,A_0^{B\rho}
 (m^2_{f_0})m_Bp_c\non \\
 &-& f_Bf_\rho\bar f_{f_0}^u\bigg[\big(\bar b_2\delta_u^p+\bar b_3
 +\bar b_{\rm 3,EW}\big)_{f_0^u \rho} +\big(\bar b_2\delta_u^p+\bar b_3
 +\bar b_{\rm 3,EW}\big)_{\rho f_0^u}\bigg] \Bigg\}, \non \\
A(\ov B^0 \to f_0\rho^0 ) &=&
i\frac{G_F}{2}\sum_{p=u,c}\lambda_p^{(d)}
 \Bigg\{ \left(a_2 \delta^p_u-a_4^p+r_\chi^\rho (a_6^p-{1\over 2}a_8^p)+{3\over 2}(a_9^p+a_7^p)
 +{1\over 2}a_{10}^p \right)_{f_0^d \rho} \non \\
 &\times& 2f_\rho F_1^{B f_0^d}(m_\rho^2)m_Bp_c
  + \Big(\bar a_2\delta_u^p+2(\bar a_3+\bar a_5)+\bar a_4+(a_6^p-{1\over
 2}a_8^p)\bar r_\chi^{f_0} \non \\
 &+& {1\over 2}(\bar a_7+\bar a_9-\bar a_{10})\Big)_{\rho f_0^u}2\bar f^u_{f_0}\,A_0^{B\rho}
 (m^2_{f_0})m_Bp_c\non \\
  &-& f_Bf_\rho\bar f_{f_0}^u\bigg[\big(\bar b_1 \delta^p_u-\bar b_3
 +{1\over 2}\bar b_{\rm 3,EW}+{3\over 2}\bar b_{\rm 4,EW}\big)_{f_0^d \rho} +\big(\bar b_1 \delta^p_u-\bar b_3
 +{1\over 2}\bar b_{\rm 3,EW}+{3\over 2}\bar b_{\rm 4,EW}\big)_{\rho f_0^d}\bigg] \Bigg\},\non \\
A(\ov B^0 \to f_0\omega ) &=&
i\frac{G_F}{2}\sum_{p=u,c}\lambda_p^{(d)}
 \Bigg\{ \left(a_2 \delta^p_u+a_4^p-r_\chi^\omega (a_6^p-{1\over 2}a_8^p)+{1\over 2}(a_9^p+a_7^p)
 -{1\over 2}a_{10}^p \right)_{f_0^d \omega} \non \\
 &\times& 2f_\omega F_1^{B f_0^d}(m_\omega^2)m_Bp_c
 - \Big(\bar a_2\delta_u^p+2(\bar a_3+\bar a_5)+\bar a_4+(a_6^p-{1\over
 2}a_8^p)\bar r_\chi^{f_0} \non \\
 &+& {1\over 2}(\bar a_7+\bar a_9-\bar a_{10})\Big)_{\omega f_0^d}2\bar
 f^d_{f_0}\,A_0^{B\omega}(m^2_{f_0})m_Bp_c \non \\
  &-& f_Bf_\omega\bar f_{f_0}^d\bigg[\big(\bar b_1 \delta^p_u+\bar b_3+2\bar b_4
 -{1\over 2}\bar b_{\rm 3,EW}+{1\over 2}\bar b_{\rm 4,EW}\big)_{f_0^d \omega} \non \\
  &+& \big(\bar b_1 \delta^p_u+\bar b_3
 +2\bar b_4-{1\over 2}\bar b_{\rm 3,EW}+ {1\over 2}\bar b_{\rm 4,EW}\big)_{\omega f_0^d}\bigg] \Bigg\},\non \\
  A(\ov B^0 \to a^+_0\rho^- ) &=&
i\frac{G_F}{\sqrt{2}}\sum_{p=u,c}\lambda_p^{(d)}
 \Bigg\{ \left( a_1\delta_u^p+ a_4^p-r_\chi^\rho (a_6^p+a_8^p)
 +a_{10}^p \right)_{a_0\rho} \non \\
 &\times& 2f_\rho F_1^{Ba_0}(m_\rho^2)m_Bp_c
 - f_Bf_\rho f_{a_0}\Big[\big(b_3+b_4
 -{1\over 2}b_{\rm 3,EW}-{1\over 2}b_{\rm 4,EW}\big)_{a_0\rho} \non
 \\  &+& \big(b_1\delta_u^p+b_4 +b_{\rm 4,EW}\big)_{\rho a_0} \Big]
 \Bigg\}, \non \\
A(\ov B^0 \to a^-_0\rho^+ ) &=&
 i\frac{G_F}{\sqrt{2}}\sum_{p=u,c}\lambda_p^{(d)}
 \Bigg\{ -\left( a_1\delta_u^p+ a_4^p+r_\chi^{a_0}( a_6^p+a_8^p)
 +a_{10}^p \right)_{\rho a_0} \non \\
 &\times& 2f_{a_0} A_0^{B\rho}(m_\rho^2)m_Bp_c
 - f_Bf_\rho f_{a_0}\Big[\big(b_3+b_4
 -{1\over 2}b_{\rm 3,EW}-{1\over 2}b_{\rm 4,EW}\big)_{\rho a_0} \non
 \\  &+& \big(b_1\delta_u^p+b_4 +b_{\rm 4,EW}\big)_{a_0\rho} \Big]
 \Bigg\}, \non \\
  A(B^- \to a^0_0\rho^- ) &=&
i\frac{G_F}{2}\sum_{p=u,c}\lambda_p^{(d)}
 \Bigg\{ \left( a_1\delta_u^p+ a_4^p-r_\chi^\rho (a_6^p+a_8^p)
 +a_{10}^p \right)_{a_0\rho} \non \\
 &\times& 2f_\rho F_1^{Ba_0}(m_\rho^2)m_Bp_c
  - \Big(\bar a_2\delta_u^p-\bar a_4-(a_6^p-{1\over
 2}a_8^p)\bar r_\chi^{a_0} \non \\
 &+& {3\over 2}(\bar a_7+\bar a_9)+{1\over 2}\bar a_{10}\Big)_{\rho a_0}2\bar
 f_{a_0}A_0^{B\rho}(m_{a_0}^2) m_Bp_c \non \\
 &-& f_Bf_\rho\bar f_{a_0}\Big[\big(\bar b_2\delta_\mu^p+\bar b_3+
 \bar b_{\rm 3,EW}\big)_{a_0\rho}-
 \big(\bar b_2\delta_\mu^p+\bar b_3 +\bar b_{\rm 3,EW}\big)_{\rho a_0} \Big]
 \Bigg\}, \non \\
  A(B^- \to a^-_0\rho^0) &=&
 i\frac{G_F}{2}\sum_{p=u,c}\lambda_p^{(d)}
 \Bigg\{ -\left( a_1\delta_u^p+ a_4^p+r_\chi^{a_0} (a_6^p+a_8^p)
 +a_{10}^p \right)_{\rho a_0} 2f_{a_0}
 A_0^{B\rho}(m_{a_0}^2)m_Bp_c \non \\
 &+& \left[a_2\delta_u^p-a_4^p+r_\chi^\rho (a_6^p-{1\over 2}a_8^p)+{1\over 2}a_{10}^p
 +{3\over 2}(a_9+a_7)\right]_{a_0\rho} 2f_{\rho} F_1^{Ba_0}(m_\rho^2)
 m_Bp_c \non \\
  &-& f_Bf_\rho f_{a_0}\Big[\big(b_2\delta_\mu^p+b_3+
 b_{\rm 3,EW}\big)_{\rho a_0}-
 \big(b_2\delta_\mu^p+b_3 +b_{\rm 3,EW}\big)_{a_0\rho} \Big]
 \Bigg\}, \non \\
  A(\ov B^0 \to a^0_0\rho^0 ) &=&
-i\frac{G_F}{2\sqrt{2}}\sum_{p=u,c}\lambda_p^{(d)}
 \Bigg\{ \left( a_2\delta_u^p- a_4^p+r_\chi^\rho (a_6^p-{1\over
 2}a_8^p) +{3\over 2}(a_9+a_7)+{1\over 2}a_{10}^p \right)_{a_0\rho} \non \\
 &\times& 2f_\rho F_1^{Ba_0}(m_\rho^2)m_Bp_c
  - \Big(\bar a_2\delta_u^p-\bar a_4-(a_6^p-{1\over
 2}a_8^p)\bar r_\chi^{a_0} \non \\
 &+& {3\over 2}(\bar a_7+\bar a_9)+{1\over 2}\bar a_{10}\Big)_{\rho a_0}2\bar
 f_{a_0}A_0^{B\rho}(m_{a_0}^2) m_Bp_c \non \\
 &+& f_Bf_\rho\bar f_{a_0}\Big[\big(\bar b_1\delta_\mu^p+\bar b_3+2\bar b_4-
 {1\over 2}(\bar b_{\rm 3,EW}-\bar b_{\rm 4,EW})\big)_{a_0\rho} \non \\
 &+& \big(\bar b_1\delta_\mu^p+\bar b_3+2\bar b_4 - {1\over 2}(\bar b_{\rm 3,EW}-\bar b_{\rm 4,EW})
 \big)_{\rho a_0} \Big]
 \Bigg\}, \non \\
  A(B^- \to a^-_0\omega) &=&
 i\frac{G_F}{2}\sum_{p=u,c}\lambda_p^{(d)}
 \Bigg\{ -\left( a_1\delta_u^p+ a_4^p+r_\chi^{a_0} (a_6^p+a_8^p)
 +a_{10}^p \right)_{\omega a_0} 2f_{a_0^-}
 A_0^{B\omega}(m_{a_0}^2)m_Bp_c \non \\
 &+& \Big[a_2\delta_u^p+2(a_3+a_5)+a_4^p-r_\chi^\omega (a_6^p-{1\over 2}a_8^p)
 \non \\
 &-& {1\over 2}a_{10}^p
 +{1\over 2}(a_9+a_7)\Big]_{a_0\omega} 2f_{\omega} F_1^{Ba_0}(m_\omega^2)
 m_Bp_c \non \\
  &-& f_Bf_\omega f_{a_0}\Big[\big(b_2\delta_\mu^p+b_3+
 b_{\rm 3,EW}\big)_{\omega a_0}+
 \big(b_2\delta_\mu^p+b_3 +b_{\rm 3,EW}\big)_{a_0\omega} \Big]
 \Bigg\}, \non \\
  A(\ov B^0 \to a^0_0\omega ) &=&
-i\frac{G_F}{2\sqrt{2}}\sum_{p=u,c}\lambda_p^{(d)}
 \Bigg\{ \Big( a_2\delta_u^p+2(a_3+a_5)+ a_4^p-r_\chi^\omega (a_6^p-{1\over
 2}a_8^p) \non \\
  &-& {1\over 2}a_{10}^p+{1\over 2}(a_9+a_7) \Big)_{a_0\omega}
 2f_\omega F_1^{Ba_0}(m_\omega^2)m_Bp_c \non \\
 &-& \Big(\bar a_2\delta_u^p+2(\bar a_3+\bar a_5)+\bar a_4+(a_6^p-{1\over
 2}a_8^p)\bar r_\chi^{a_0} \non \\
 &+& {1\over 2}(\bar a_7+\bar a_9-\bar a_{10})\Big)_{\omega a_0}2\bar
 f_{a_0}A_0^{B\omega}(m_{a_0}^2)m_Bp_c \non \\
 &-& f_Bf_\omega \bar f_{a_0}\Big[\big(-\bar b_1\delta_\mu^p+\bar b_3-
 {1\over 2}\bar b_{\rm 3,EW}-{3\over 2}\bar b_{\rm 4,EW}\big)_{a_0\omega} \non \\
 &+& \big(-\bar b_1\delta_\mu^p+\bar b_3 - {1\over 2}\bar b_{\rm 3,EW}-{3\over 2}\bar b_{\rm 4,EW}
 \big)_{\omega a_0} \Big]
 \Bigg\}, \non \\
A(B^- \to K^{*-}_0\phi ) &=&
i\frac{G_F}{\sqrt{2}}\sum_{p=u,c}\lambda_p^{(s)}
 \Bigg\{ \left( a_3+a_4^p+a_5-r_\chi^\phi(a_6^p-{1\over 2}a_8^p)-{1\over 2}(a_7+a_9+a_{10}^p)\right)_{ K^*_0\phi} \non \\
 &\times& 2f_\phi F_1^{BK^*_0}(m_\phi^2)m_Bp_c
 - f_Bf_\phi f_{K^*_0}\big(b_2\delta_u^p+b_3
 +b_{\rm 3,EW}\big)_{K^*_0\phi} \Bigg\}, \non \\
A(\ov B^0 \to \ov K^{*0}_0\phi ) &=&
i\frac{G_F}{\sqrt{2}}\sum_{p=u,c}\lambda_p^{(s)}
 \Bigg\{ \left( a_3+a_4^p+a_5-r_\chi^\phi(a_6^p-{1\over 2}a_8^p)-{1\over 2}(a_7+a_9+a_{10}^p)\right)_{ K^*_0\phi} \non \\
 &\times& 2f_\phi F_1^{BK^*_0}(m_\phi^2)m_Bp_c
 - f_Bf_\phi f_{K^*_0}\big(b_3
 -{1\over 2}b_{\rm 3,EW}\big)_{K^*_0\phi} \Bigg\}, \non \\
A(B^- \to \ov K^{*0}_0\rho^- ) &=&
i\frac{G_F}{\sqrt{2}}\sum_{p=u,c}\lambda_p^{(s)}
 \Bigg\{ -\left( a_4^p+r_\chi^{K^*_0}(a_6^p-{1\over 2}a_8^p)
 -{1\over 2}a_{10}^p\right)_{\rho K^*_0} \non \\
 &\times& 2f_{K_0^*}A_0^{B\rho}(m_{K_0^*}^2)m_Bp_c
 - f_Bf_\rho f_{K^*_0}\big(b_2\delta_u^p+b_3
 +b_{\rm 3,EW}\big)_{\rho K^*_0} \Bigg\}, \non \\
A(B^- \to K^{*-}_0\rho^0 ) &=&
i\frac{G_F}{2}\sum_{p=u,c}\lambda_p^{(s)}
 \Bigg\{-\left( a_1\delta_u^p+a_4^p+r_\chi^{K^*_0}(a_6^p+a_8^p)
 +a_{10}^p \right)_{\rho K^*_0} \non \\
 &\times& 2f_{K_0^*}A_0^{B\rho}(m_{K_0^*}^2)m_Bp_c+\left[a_2\delta_u^p+{3\over
 2}(a_9+a_7)\right]_{K^*_0\rho}2f_\rho F_1^{BK^*_0}(m_\rho^2)m_Bp_c
 \non \\
 &-& f_Bf_\rho f_{K^*_0}\big(b_2\delta_u^p+b_3
 +b_{\rm 3,EW}\big)_{\rho K^*_0} \Bigg\}, \non \\
 A(\ov B^0 \to K^{*-}_0\rho^+ ) &=&
i\frac{G_F}{\sqrt{2}}\sum_{p=u,c}\lambda_p^{(s)}
 \Bigg\{ -\left( a_1\delta_u^p+ a_4^p+r_\chi^{K^*_0}a_6^p
 +a_{10}^p+r_\chi^{K^*_0}a_8^p \right)_{\rho K^*_0} \non \\
 &\times&
 2f_{K_0^*}A_0^{B\rho}(m_{K_0^*}^2)m_Bp_c
 - f_Bf_\rho f_{K^*_0}\big(b_3
 -{1\over 2}b_{\rm 3,EW}\big)_{\rho K^*_0} \Bigg\}, \non \\
 A(\ov B^0 \to \ov K^{*0}_0\rho^0 ) &=&
i\frac{G_F}{2}\sum_{p=u,c}\lambda_p^{(s)}
 \Bigg\{ -\left( -a_4^p-r_\chi^{K^*_0}(a_6^p-{1\over 2}a_8^p)
 +{1\over 2}a_{10}^p \right)_{\rho K^*_0} \non \\
 &\times&
 2f_{K_0^*}A_0^{B\rho}(m_{K_0^*}^2)m_Bp_c+\left[a_2\delta_u^p+{3\over
 2}(a_9+a_7)\right]_{K^*_0\rho}2f_\rho F_1^{BK^*_0}(m_\rho^2)m_Bp_c
 \non \\
 &-& f_Bf_\rho f_{K^*_0}\big(-b_3
 +{1\over 2}b_{\rm 3,EW}\big)_{\rho K^*_0} \Bigg\}, \non \\
A(B^- \to K^{*-}_0\omega ) &=&
i\frac{G_F}{2}\sum_{p=u,c}\lambda_p^{(s)}
 \Bigg\{ \left[a_2\delta_u^p+2(a_3+a_5)+{1\over
 2}(a_9+a_7)\right]_{K^*_0\omega}2f_\omega F_1^{BK^*_0}(m_\omega^2)m_Bp_c
 \non \\
  &-& \left( a_1\delta_u^p+a_4^p+r_\chi^{K^*_0}(a_6^p+a_8^p)
 +a_{10}^p \right)_{\omega K^*_0}
 2f_{K_0^*}A_0^{B\omega}(m_{K_0^*}^2)m_Bp_c \non \\
 &-& f_Bf_\omega f_{K^*_0}\big(b_2\delta_u^p+b_3
 +b_{\rm 3,EW}\big)_{\omega K^*_0} \Bigg\}, \non \\
 A(\ov B^0 \to \ov K^{*0}_0\omega ) &=&
i\frac{G_F}{2}\sum_{p=u,c}\lambda_p^{(s)}
 \Bigg\{ \left[a_2\delta_u^p+2(a_3+a_5)+{1\over
 2}(a_9+a_7)\right]_{K^*_0\omega}2f_\omega F_1^{BK^*_0}(m_\omega^2)m_Bp_c
 \non \\
 &-& \left(a_4^p+r_\chi^{K^*_0}(a_6^p-{1\over 2}a_8^p)
 -{1\over 2}a_{10}^p \right)_{\omega K^*_0}
 2f_{K_0^*}A_0^{B\rho}(m_{K_0^*}^2)m_Bp_c \non \\
 &-& f_Bf_\omega f_{K^*_0}\big(b_3
 -{1\over 2}b_{\rm 3,EW}\big)_{\omega K^*_0} \Bigg\},
 \en
where  $\lambda_p^{(q)}\equiv V_{pb}V_{pq}^*$ with $q=d,s$ and
 \be \label{eq:r}
 &&  \bar r_\chi^{f_0}(\mu)={2m_{f_0}\over
 m_b(\mu)}, \quad \bar r_\chi^{a_0^0}(\mu)={2m_{a_0^0}\over
 m_b(\mu)}, \quad
   r^{a_0^\pm}_\chi(\mu)={2m_{a_0^\pm}^2\over
 m_b(\mu)(m_d(\mu)-m_u(\mu))},
 \en
In Eq. (\ref{eq:SDAmp}), we encounter terms such as $a_i f_{f_0}$,
which appears to vanish at first sight as $f_{f_0}=0$. However,
when $f_{f_0}$ combines with $\mu_{f_0}$ appearing in the twist-2
LCDA of the scalar meson [see Eq. (\ref{eq:twist2wf})], it becomes
finite, namely, $f_{f_0}\,\mu_{f_0}=\bar f_{f_0}$. Therefore, the
effective Wilson coefficients $\bar a_i$ in Eq. (\ref{eq:SDAmp})
are defined as $a_i \mu^{-1}_S$  and they can be obtained from Eq.
(\ref{eq:ai}) by retaining only those terms that are proportional
to $\mu_S$. Specifically,
 \be \label{eq:barai}
 \bar a_i^p(M_1M_2) ={c_{i\pm1}\over N_c}\,{C_F\alpha_s\over
 4\pi}\Big[\bar V_i(M_2)+{4\pi^2\over N_c}\bar H_i(M_1M_2)\Big]+\bar
 P_i^p(M_2).
 \en
The LCDA of the neutral scalar meson in the bar quantities, $\bar
V_i(S)$, $\bar P_i(S)$ and $\bar H_i(M1,M2)$ is replaced by
$\bar\Phi_S$ which has the similar expression as Eq.
(\ref{eq:twist2wf}) except that the first constant term does not
contribute and the term $f_S\,\mu_S$ is factored out
 \be
 \bar \Phi_S(x,\mu)= 6x(1-x)\sum_{m=1}^\infty
 B_m(\mu)\,C_m^{3/2}(2x-1).
 \en
In Eq. (\ref{eq:barai}),
 \be
 \bar V_i(S)=\cases{\left({11\over 2}-3i\pi\right)B_1^S+\left({79\over
36}-{2i\pi\over 3}\right)B_3^S+\cdots; & for~$i=1-4,9,10$, \cr
-\left({11\over 2}-3i\pi\right)B_1^S-\left({79\over
36}-{2i\pi\over 3}\right) B_3^S+\cdots; & for~$i=5,7$, \cr 0; &
for~$i=6,8$.}
 \en
The annihilation terms $\bar b_i$ have the same expressions as Eq.
(\ref{eq:bi}) with $r_\chi^S$ and $\mu_S B_i$ replaced by $\bar
r_\chi^S$ and $B_i$, respectively.

For the CKM matrix elements, we use the Wolfenstein parameters
$A=0.806$, $\lambda=0.22717$, $\bar \rho=0.195$ and $\bar
\eta=0.326$ \cite{CKMfitter}.  For the running quark masses we
shall use
 \be \label{eq:quarkmass}
 && m_b(m_b)=4.2\,{\rm GeV}, \qquad~~~ m_b(2.1\,{\rm GeV})=4.95\,{\rm
 GeV}, \qquad m_b(1\,{\rm GeV})=6.89\,{\rm
 GeV}, \non \\
 && m_c(m_b)=1.3\,{\rm GeV}, \qquad~~~ m_c(2.1\,{\rm GeV})=1.51\,{\rm
 GeV}, \non \\
 && m_s(2.1\,{\rm GeV})=90\,{\rm MeV}, \quad m_s(1\,{\rm GeV})=119\,{\rm
 MeV}, \non\\
 && m_d(1\,{\rm GeV})=6.3\,{\rm  MeV}, \quad~ m_u(1\,{\rm GeV})=3.5\,{\rm
 MeV}.
 \en
The uncertainty of the strange quark mass is specified as
$m_s(2.1\,{\rm GeV})=90\pm20$ MeV.


\end{document}